\documentclass[aps,prd,twocolumn,nofootinbib,showpacs,preprintnumbers]{revtex4}

\usepackage{epsfig}
\usepackage{amsmath}
\usepackage{amssymb}
\usepackage[dvipdfm]{hyperref} 
\def\sn{\,{\rm sn}}
\def\e{e}
\def\d{d}
\def\i{i}
\def\K{K}
\def\C{C}
\def\D{{\cal D}}
\def\Tau{{\cal T}}
\def\x{r}
\def\epst{\zeta}
\def\epss{\varepsilon}
\newcommand{\rf}[1]{(\ref{#1})}
\newcommand{\eq}[1]{Eq.~(\ref{#1})}

\def\be{\begin{equation}}
\def\ee{\end{equation}}
\def\bea{\begin{eqnarray}}
\def\eea{\end{eqnarray}}
\def\LA{\left\langle}
\def\RA{\right\rangle}
\newcommand{\non}{\nonumber \\*}

\newcommand{\pint}{\int\hspace{-1.17em}\not\hspace{.75em}}

\begin{document}

\preprint{ITEP--TH--49/10}

\title{Effective String Theory and QCD Scattering Amplitudes}

\author{Yuri Makeenko}
\altaffiliation[Also at]{ the Institute for Advanced Cycling,
Blegdamsvej 19, 2100 Copenhagen \O, Denmark}

\affiliation{Institute of Theoretical and Experimental Physics\\
B.~Cheremushkinskaya 25, 117218 Moscow, Russia}
\email{makeenko@itep.ru\ \ \ makeenko@nbi.dk}

\date{December 22, 2010}

\begin{abstract}
QCD string is formed at the distances larger than the confinement scale and 
can be described by the Polchinski--Strominger effective string theory
with a nonpolynomial action, which has nevertheless a well-defined 
semiclassical expansion around a long-string ground state.
We utilize modern ideas about the Wilson-loop/scattering-amplitude duality 
to calculate scattering amplitudes and show that the expansion parameter 
in the effective string theory is small in the Regge kinematical regime.
For the amplitudes we obtain the Regge behavior with a linear trajectory of 
the intercept $(d-2)/24$ in $d$ dimensions, which is computed semiclassically 
as a momentum-space L\"uscher term, and discuss an application to meson 
scattering amplitudes in QCD.

\end{abstract}


\maketitle

\section{Introduction}

Recently there has renewed interest in the long-standing problem of the 
relation between strings and QCD. On the one hand, some properties 
of Wilson loops have been understood 
via the AdS/CFT correspondence~\cite{Mal98}, where
the Wilson loop in ${\cal N}=4$ super Yang--Mills at large coupling constant
is described by the supergravity approximation to an open superstring of type 
IIB on $AdS_5\times S^5$ background, whose worldsheet is 
a minimal surface extended to the 5th dimension. This approach has resulted
in numerous applications of holographic duals to QCD. 
On the other hand, the lattice QCD simulations indicate
(see the review~\cite{Tep09}, references therein
and the subsequent paper~\cite{ABT10}) 
that the Nambu--Goto string very well approximates 
the QCD string for a wide range of distances.  

An old result~\cite{MM79} 
is that the Nambu--Goto string is not an exact solution to the loop equation 
of large-$N$ QCD, but rather its asymptote --- the area law --- {\em is}\/ 
a self-consistent solution for asymptotically large loops.
Extra degrees of freedom, populating the string worldsheet, are required 
to reproduce a factorized structure on the right-hand side 
of the loop equation at intermediate distances and/or a proper behavior
of Wilson loops for the case of selfintersections. 
These degrees become frozen for large loops = long strings (in the units
of the QCD confinement scale), that makes it possible
to perform an expansion in the inverse area of the minimal surface, 
spanned by the loop, which has the meaning of a semiclassical expansion.
This leads us to an ideology of an effective QCD string, formed by fluxes 
of the Yang--Mills field, which is 
consistent~\cite{Ole85} at large distances.

A beautiful example of how such an effective string theory works is 
a closed string winding along 
a compact direction of a large radius $R$. It is described by 
a nonpolynomial action~\cite{PS91}
\begin{equation}
S_{\rm eff}=2\K \int \d^2 z \,\partial X \cdot \bar \partial X + 
\frac{d-26}{24\pi}  \int \d^2 z \, 
\frac{\partial^2 X \cdot  \bar \partial^2 X}{\partial X \cdot \bar \partial X}
+\ldots, 
\label{effaction}
\end{equation} 
where the conformal anomaly is expressed (modulo total derivatives and 
the constraints) via an induced metric 
\be
\e^{\varphi_{\rm ind} }=2 \,\partial X \cdot \bar \partial X 
\label{indmetric}
\ee
in the conformal gauge, which is not treated
independently as distinct from the Polyakov formulation. 
This effective string theory
has been analyzed using the conformal field theory technique order
by order in $1/R$~\cite{PS91,Dru04}, 
revealing the spectrum~\cite{Arv83} of the
Nambu--Goto string in $d$-dimensions. 

The goal of this Paper is to expand the effective string theory approach to
calculations of QCD meson scattering amplitudes in the Regge kinematical regime,
where a semiclassical expansion is applicable as will be momentarily 
explained.
These scattering amplitudes are represented in the large-$N$ limit (or
in the quenched approximation) as sums over paths of the Wilson loops.
Remarkably, large loops dominate the sum over paths in the Regge
kinematical regime when Mandelstam's variables $s$ is large and $t$ is fixed,
as it has been shown in Ref.~\cite{MO08}, so an effective 
string theory ideology is then applicable. 

In obtaining this result,
it was crucial to use the manifestly reparametrization-in\-var\-i\-ant
representation~\cite{Pol97} of large Wilson loops
in the form of the path integral over repa\-ram\-e\-trizaions of the 
boundary contour $x^\mu(t)$: 
\begin{equation}
W[x(\cdot)]= \int {\cal D}_{\rm diff} t(s)\e^{-\K S[x(t)]},
\label{ansatz}
\end{equation}
where $\K=1/2\pi\alpha'$ is the string tension  and
\begin{equation}
S[x(t)]=\frac{1}{4\pi}
\int_{-\infty}^{+\infty}\frac{\d s_1 \d s_2}{(s_1-s_2)^2} \,
{\left[x(t(s_1))-x(t(s_2))\right]^2}.
\label{Di}
\end{equation}
We have used the notation $W[x(\cdot)]$ on the left-hand side of \eq{ansatz}
to emphasize its reparametrization invariance.

The functional (\ref{Di}) is known in mathematics as the Douglas integral 
\cite{Dou31}, whose minimum with respect to 
reparametrizing functions $t(s)$ coincides with the minimal area.
The path integral in \eq{ansatz} is
thus dominated for large loops by a saddle point, 
giving the area law. This remarkably holds for loops of an arbitrary shape,
even not necessarily planar.
The area-law behavior of the reparametrization path integral is 
of course associated with a classical string.
As is shown in Ref.~\cite{MO10a}, quantum
fluctuations around the saddle point reproduce in the quadratic 
approximation the L\"uscher term in $d=26$ dimensions, which is
usually associated with quantum fluctuations around the minimal
surface in the semiclassical approximation. This is perhaps not
surprising because the ansatz~\rf{ansatz} emerges as the Dirichlet 
disk amplitude for the Polyakov string in the critical dimension $d=26$. 

The fact that $S[x(t)]$ in \eq{ansatz} is quadratic in $x^\mu(t)$ makes it
possible to perform its Fourier transformation by doing the Gaussian
path integral over $x^\mu(t)$, which results in the scattering
amplitude again of the type of the right-hand side of \eq{ansatz} with
$x^\mu(t)$ substituted by the function $p^\mu(t)/\K$, where $p^\mu(t)$ is 
a step function, whose discontinuities are momenta of colliding particles. 
Thus all nonlinearities are hidden in the reparametrization path
integral which can be partially done, while the rest is represented as
an integral over the Koba--Nielsen variables known from dual 
resonance models. 
This path integral over reparametrizations goes over 
{\em subordinated}\/ functions 
(i.e.\ those having $\d t(s)/\d s\geq 0$) with a certain measure which
respects reparametrization invariance and whose properties are considered 
in some detail in Refs.~\cite{MO08,BM09}. What is most important is
that the resulting scattering amplitudes~\cite{MO08,MO10b} possess 
projective invariance and are consistent off-shell. 
Once again, this is intimately related to the presence of the reparametrization
path integral in \eq{ansatz}, which factorizes in the on-shell scattering
amplitudes of the fundamental string 
(i.e.\ for tachyonic scalars, massless vectors etc.), reproducing
the usual Koba--Nielsen amplitude, and correspondingly plays no role then. 
On the contrary, excitations of QCD string should reproduce the meson spectrum,
i.e.\ the vector state is a massive $\rho$-meson, which explains why  
scattering amplitudes are required off-shell. 
Off-shell amplitudes of this kind were previously 
obtained~\cite{DiV86} (see the review~\cite{DiV92} and the subsequent 
papers~\cite{CLP98}) for the Polyakov quantization of the critical string, 
using the Lovelace choice~\cite{Lov70} of the $N$-Reggeon vertex 
instead of the usual vertex operator.
However, their extension to $d=4$ dimensions is still missing to my knowledge.

In the present Paper we derive scattering amplitudes for a noncritical 
effective open string theory with the action~\rf{effaction} in 
the semiclassical approximation justified by the Regge kinematical regime,
where the expansion parameter $1/{\ln(s/t)}$ is small.  
The technique used is pretty much in the spirit of the
Wilson-loop/scattering-amplitude duality recently elaborated~\cite{AM07a,DSK07}
(for a review see Ref.~\cite{AR08}) for ${\cal N}=4$ super Yang--Mills. 
The calculation is analogous to that of the L\"uscher term for a rectangle,
except it is performed in momentum space.
As a result, we obtain 
the Regge behavior of scattering amplitudes with a linear trajectory
\be
\alpha(t)=\frac{d-2}{24} +\alpha' t.
\label{linea}
\ee
We then discuss an application of this result to large-$N$ QCD, 
where meson scattering amplitudes are represented as sums over paths of
the Wilson loop.
We demonstrate that large loops dominate the sum over paths in the Regge
kinematical regime of large $s$ and fixed $-t$, 
so the effective string theory representation 
of the Wilson loop is expected to work. Alternatively, perturbative QCD
is expected to work when both $s$ and $-t$ are large. 
We also discuss how a linear Regge trajectory
of the type in \eq{linea} appears for spinor quarks.

\section{The classical limit}

\subsection{Review of Douglas' minimization\label{ss:Douglas}}

Let us consider \eq{ansatz} as a representation of
the disk amplitude for bosonic string with the Dirichlet boundary condition 
in $d= 26$ dimensions. As is already mentioned in the Introduction,
this representation can be derived in the Polyakov formulation by
integrating over $X^\mu(x,y)$ in the bulk with the boundary condition
\be
X^\mu(s,0)= x^\mu (t(s))\,.
\label{b.c.}
\ee
The form of the boundary action~\rf{Di} depend on the choice of
coordinates parametrizing the world sheet, as the Green function 
of the Laplace operator does. Equation~\rf{Di} is written for the
upper half-plane (UHP): $z=x+i y\in\hbox{UHP}$, bounded by the real axis.
While the Polyakov action is invariant under conformal transformations,
they change, in general, the shape of the boundary, so the Douglas
integral~\rf{Di} changes accordingly. The only conformal transformation
that maps UHP onto itself is $SL(2;\Bbb{R})$, which results in the
projective transformation at the boundary:
\be
s\longrightarrow \frac{as+b}{cs+d},\qquad ad-bc=1.
\label{project}
\ee
The Douglas integral~\rf{Di} is invariant under it. 

It is instructive to compare the UHP parametrization with
a more physical parametrization through worldsheet coordinates,
which take values in a rectangle and are usually associated 
with propagation of an open string of the length $R$ during the time $T$.
These two coordinate choices are related by the Schwarz--Christoffel mapping,
which will be extensively used below when calculating 
a semiclassical correction. For the purposes of the present Paper, the
former parametrization has some advantages over the latter.
Firstly, the Green function in \eq{Di} looks
simpler for UHP than for a rectangle.%
\footnote{The Dirichlet Green functions for UHP and a rectangle
are displayed below in Sect.~\ref{sec:3c}
(see Eqs.~\rf{56} and \rf{Greenomega}).} 
Secondly, after the decomposition
\be
X^\mu =X^\mu_{\rm cl}+ Y^\mu_{\rm q},
\label{deco}
\ee
where $X^\mu_{\rm cl}$ obeys the Laplace equation and the boundary condition,
so that $Y^\mu_{\rm q}=0$ at the boundary, the path integral over $Y^\mu_{\rm q}$
does {\em not}\/ depend on the boundary contour $x^\mu$ for the UHP
parametrization. This is the reason why the boundary path integral in 
\eq{ansatz} captures fluctuations of the critical string around 
the minimal surface, as was explicitly demonstrated in Ref.~\cite{MO10a}.
This is in contrast to the parametrization by a rectangle, when semiclassical
stringy fluctuations resides in a determinant coming from the path integral 
over $Y^\mu_{\rm q}$ as is well-known. 
We shall return to this issue in Sect.~\ref{s:Lu}.

The minimum of the Douglas integral is reached for the function $t(s)$
obeying
\be
\pint \d t_1 \frac{\dot x(t)\cdot \dot x(t_1)}{\left[s(t)-s(t_1)\right]}=0,
\label{Doumin}
\ee
where $s(t)$ denotes inverse to $t(s)$. The minimal surface can then be
reconstructed in UHP from its boundary value $x^\mu(t(s))$ by the Poisson 
formula
\bea
X^\mu(x,y)&=&\int_{-\infty}^{+\infty} \frac {\d s}{\pi} \,
\frac {x^\mu (t(s))\,y}{(x-s)^2+y^2}\non &=&
\int_{-\infty}^{+\infty} \frac {\d t}{\pi} \,
\dot x^\mu (t)\arctan \frac{x-s(t)}y.
\label{harmox}
\eea
This function is obviously harmonic in UHP and satisfies \eq{b.c.}.
The presence of the reparametrizing function $t(s)$ guarantees that \rf{harmox} 
obeys the conformal gauge if $t(s)=t_*(s)$ with $t_*(s)$ being inverse to 
the minimizing function $s_*(t)$. This is demonstrated in Appendix~\ref{appA}.
While Douglas' theorem was originally proven for Euclidean space,
the consideration of Appendix~\ref{appA} shows that it applies for
a space-like surface in Minkowski space as well.  
The necessity of a reparametrization of the boundary for 
consistency with the conformal gauge in the Polyakov formulation 
of an open string was pointed out in Ref.~\cite{Pol81}. 


\subsection{Polygonal loop with $(x_{i+1}-x_i)=\Delta p_i /\K$}

Since the boundary action~\rf{Di} is quadratic in $x^\mu(t)$, the
functional Fourier transformation of \eq{ansatz} 
to momentum space equals~\cite{MO08} 
\begin{equation}
A[p(\cdot)]\equiv\int \D x^\mu\,\e^{\i \int \d t\,p \cdot \dot x}\,W[x(\cdot)] =
W \left[p(\cdot)/K\right],
\label{msWL}
\end{equation}
which looks exactly like
the right-hand side of \eq{ansatz} with $x^\mu(t)$ substituted by
the trajectory
\be
x^\mu (t)= \frac{1}\K p^\mu(t).
\label{xp}
\ee 
For piecewise constant $p^\mu(t)$ this disk amplitude
is proportional to the scattering amplitude.
We shall make use of this remarkable fact applying the technique, 
developed for a noncritical string with Dirichlet boundary conditions, 
to semiclassical calculations of the scattering amplitudes.

Substituting in \eq{msWL} the smeared stepwise
\bea
p^\mu(t)&=&\frac 1\pi\sum_i \Delta p_i^\mu \arctan \frac{(t-t_i)}{\epst_i}
\non &{\rightarrow}&
\frac1{2} \sum_i \Delta p_i^\mu \,{\rm sign}\,(t-t_i)
\label{stepx}
\eea
for $(t_i-t_{i-1})\gg \epst_i,\epst_{i-1}$, that results in polygonal $x^\mu(t)$,
we have for the amplitude explicitly
\be
A\left(\{\Delta p_i\}\right)= W \left[p(\cdot)/\K\right]
\label{Psip}
\ee
with $p^\mu(t)$ given by \eq{stepx}. The discontinuities $\Delta p_i^\mu$ of 
$p^\mu(t)$ are the particle momenta.

Let us calculate the minimal area for such nonplanar contours. 
Since we are interested in the Regge limit of
$s\gg- t \gg -\Delta p_i^2$, we can set $\Delta p_i^2=0$ to have light-like
edges like in Refs.~\cite{AM07a,DSK07}.
The case of $\Delta p_i^2\neq 0$ will be considered in Sect.~\ref{s:neq}.

The Douglas integral then reads
\begin{eqnarray}
\K S&=&-\alpha'\int\d t\,\d t'\,\dot p(t)\cdot\dot p(t')\ln|s(t)-s(t')| 
\non &=&-\alpha'
\sum_{i,j\neq i}  \Delta p_i \cdot \Delta p_j \ln|s_i-s_j| 
\label{no1}
\end{eqnarray}
with $s_i=s(t_i)$. Here the values $t_i$'s, at which $p^\mu(t)$ has (smeared)
discontinuities, are fixed by the initial parametrization, while
the Douglas minimization is to be performed with respect to $s_i$.
Nothing depends on $s(t)$ at the intermediate points $t\in (t_{i-1},t_i)$,
which is a zero mode as is explained in Ref.~\cite{MO08}. 

The Douglas minimization equation~\rf{Doumin}
is trivially satisfied for the given polygonal $x^\mu(t)$
at the intermediate points, when $t$ is not 
close to $t_i$'s, because then $\dot x^\mu(t)=0$. 
For $t=t_i$ we rewrite \eq{Doumin} as
\be
\sum_{j\neq i} \frac{\Delta p_i \cdot \Delta p_j}{s_i-s_j}=0.
\label{DouminM}
\ee
Only $M-3$ of these $M$ equations are independent because of the 
invariance under the projective transformation of $s_i$'s.
Thus the Douglas minimization determines only $M-3$ values of $s_i$'s,
while three of them remain arbitrary. The minimal surface does not
depend on these three values.

For $M=4$ we obtain from \eq{DouminM}
\be
s_2\mbox{}\,_*= 
s_1+\frac{s s_{41} s_{31}}{s s_{41}+t s_{43}}=
s_3-\frac{t s_{43} s_{31}}{s s_{41}+t s_{43}}
\label{s2*gen}
\ee
with arbitrary $s_1$, $s_3$ and $s_4$.
In the usual way we can set $s_1=0$, $s_3=1$,
$s_4=\infty$, after which the solution~\rf{s2*gen} simplifies to
\be
s_2\mbox{}\,_*=\frac{s}{s+t}.
\label{s2*}
\ee 
This is nothing but the well-known saddle point of the Veneziano 
amplitude at large $-s$ and $-t$. 

At the minimum we shall get the minimal area
\be
\K S_{\rm min} = \alpha' s \ln \frac s{s+t}+\alpha' t \ln \frac t{s+t}
\stackrel{s\gg t}\to -\alpha' t \ln \frac st
\label{Smin}
\ee
whose exponential reproduces the classical Regge behavior of the
scattering amplitude:
\be
A(s,t)= \e^{-\K S_{\rm min}} \propto s^{\alpha' t}.
\label{Reggestick}
\ee 

\subsection{Reconstruction of the minimal surface}

For polygonal $x^\mu(t)$ given by Eqs.~\rf{xp}, \rf{stepx} we can 
reconstruct the harmonic function in UHP by the Poisson
formula \rf{harmox} which satisfies the boundary condition~\rf{b.c.}.

From Eqs.~\rf{harmox}, \rf{stepx} we have
\be
X^\mu(x,y)=  \frac 1{\pi\K}\sum_i \Delta p_i^\mu  
\arctan \frac{(x-s_i)}{y+\epss_i}.
\label{amostX}
\ee
It is instructive to see how the boundary contour~\rf{stepx} 
is reproduced by this formula for $y=0$. 
For $t\approx t_i$ we have 
\be
\frac{s(t)-s(t_i)}{\varepsilon_i} \to \frac{s'(t_i)(t-t_i)}{\varepsilon_i}
= \frac{(t-t_i)}{\epst_i},
\ee
where $\epss_i= s'(t_i)\epst_i $ in accordance with the reparametrization
covariance.
As $\epst_i\to0$ we reproduce the step function~\rf{stepx}
which results in the harmonic function~\rf{amostX} with $\epss_i=0$.
It is used below in this Subsection because there are no
divergences in the $\epss_i\to 0$ limit at the classical level. 

The domain of both $s<0$ and $t<0$ corresponds to scattering in the 
$u$-channel:
\bea
\Delta p_1^\mu &=& \left( E,p,0,0 \right), \non
\Delta p_2^\mu &=& \left( -E,-p \cos \theta,-p \sin \theta,0 \right), \non
\Delta p_3^\mu &=& \left( E,- p  ,0,0 \right), \non
\Delta p_4^\mu &=& \left( -E,p \cos \theta ,p \sin \theta,0 \right) ,
\label{centmass}
\eea
where
\be
\cos \theta = \frac t{s+t},
\qquad (1-\cos\theta)= \frac s{s+t}.
\ee
From \eq{s2*gen} we then have
\be
\cos \theta =\frac{s_{32}s_{41}}{s_{42}s_{31}},
\qquad (1-\cos\theta)=\frac{s_{21}s_{43}}{s_{42}s_{31}}.
\ee

\begin{figure}
\vspace*{3mm}
\includegraphics[width=8cm]{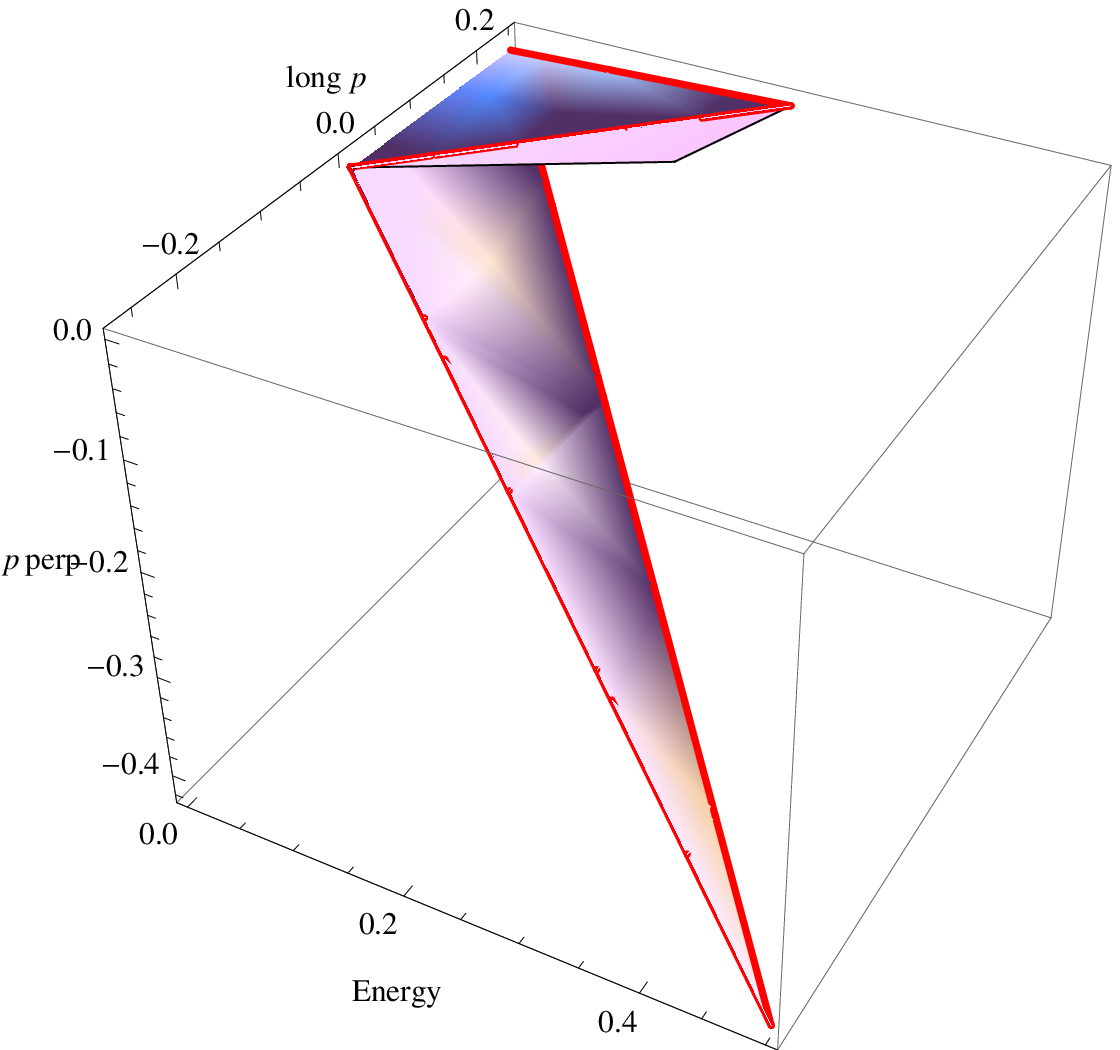} 
\includegraphics[width=8cm]{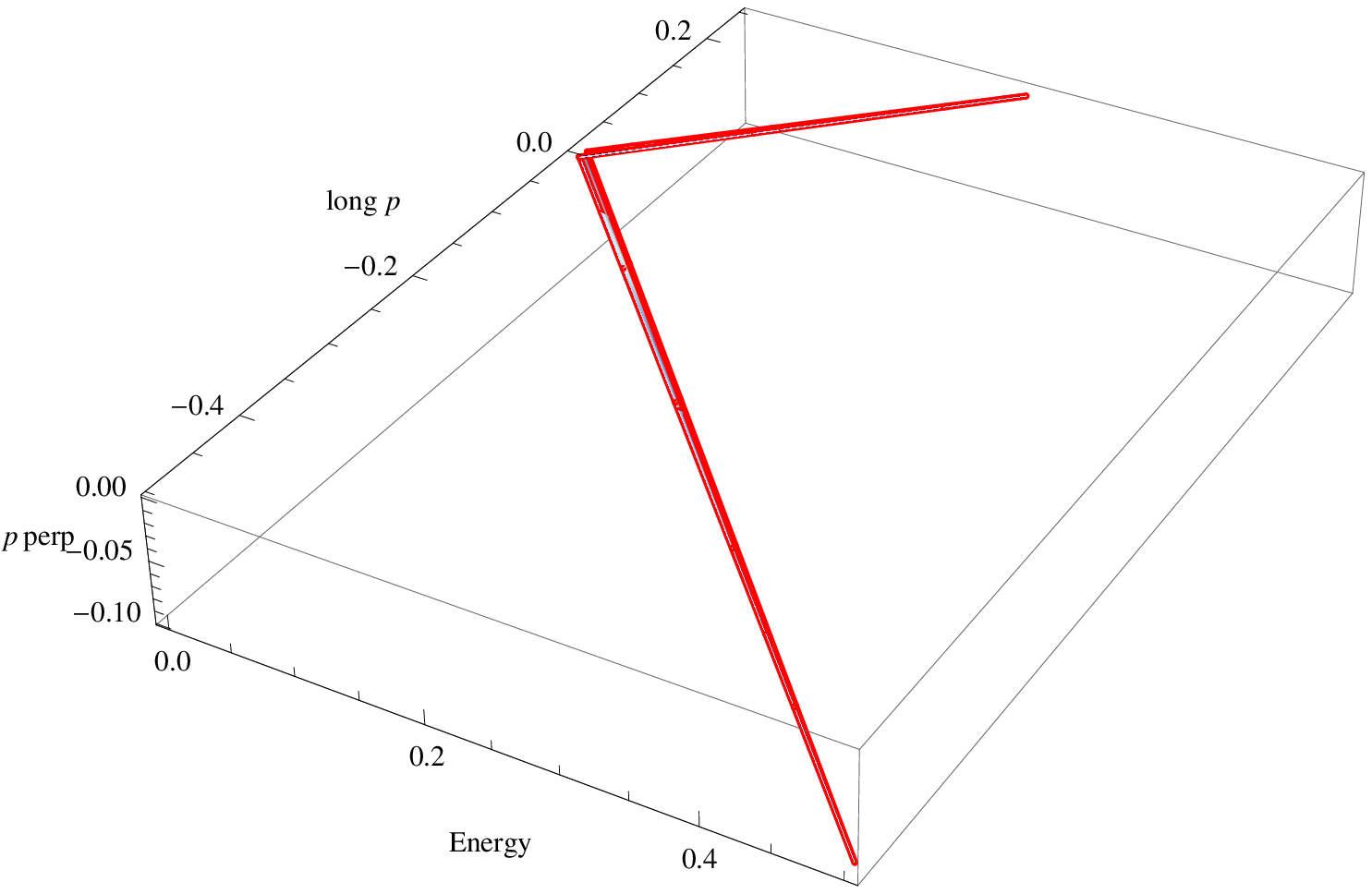} 
\caption[]{Minimal surface spanned by the contour~\rf{stepx}
for $\theta=1.0$ (left) and $\theta=0.2$ (right).}   
\label{fi:minsurf}
\end{figure}

The minimal surface spanned by the contour~\rf{stepx} with $\Delta p_i$'s 
given by \eq{centmass} is depicted
in Fig.~\ref{fi:minsurf} for $\theta=1.0$ and $\theta=0.2$.
With decreasing the scattering angle $\theta$, we move from the one in
the left figure to the one in the right figure with decreasing the
minimal area which tends to $0$ as $\theta\to0$.
It is a space-like surface embedded in Minkowski space.
One-loop divergences, associated with its transverse fluctuations, 
will be regularized by setting $\epss_i\neq0$, as is described below.  

The induced metric $g_{ab}=\partial_a X \cdot \partial_b X$ of 
the minimal surface, spanned the polygon given by \eq{stepx}, reads
\be
g_{12}=g_{21}=-\frac1{\pi^2\K^2} \sum_{i,j\neq i} 
\frac{\Delta p_i \cdot \Delta p_j\, (x-s_i)y}{[(x-s_i)^2+y^2][(x-s_j)^2+y^2]}
\label{g12}
\ee
and
\be
g_{11}=\frac1{\pi^2\K^2} \sum_{i,j\neq i} 
\frac{\Delta p_i \cdot \Delta p_j\, y^2}{[(x-s_i)^2+y^2][(x-s_j)^2+y^2]},
\label{g11} 
\ee
\be
g_{22}=\frac1{\pi^2\K^2} \sum_{i,j\neq i} 
\frac{\Delta p_i \cdot \Delta p_j\, (x-s_i)(x-s_j)}
{[(x-s_i)^2+y^2][(x-s_j)^2+y^2]}.
\label{g22}
\ee
Using the identities of Appendix~A, it can be shown that $g_{12}$,
given by \eq{g12}, vanishes and $g_{11}$, given by \eq{g11}, coincides with 
$g_{22}$, given by \eq{g22}, if \eq{DouminM} is satisfied.
Then the induced metric is conformal:
\be
g_{ab} =\e^{\varphi} \delta_{ab}
\ee
with
\be
\e^{\varphi(x,y)}=\frac1{\pi^2\K^2} \sum_{i,j\neq i} 
\frac{\Delta p_i \cdot \Delta p_j\, y^2}{[(x-s_i)^2+y^2][(x-s_j)^2+y^2]}.
\label{mmm}
\ee
When $y\to0$, this function vanishes except in the vicinities of $s_i$'s.
This implies that the boundary metric vanishes in the corners 
of the polygon and may nonvanish only along the edges.

When $M=4$ and the projective symmetry is not fixed, 
\eq{s2*gen} holds and \eq{s2*} changes to
\be
\frac{s_{21} s_{43}}{s_{31} s_{42}}=\frac{s}{s+t}.
\ee
We then get for the induced metric
\begin{equation}
\e^{\varphi(x,y)}=
-\frac{st s^2 _{42}  s^2_{31} y^2}
{\pi^2\K^2(s+t)\prod_{i=1}^4 [(x-s_i)^2+y^2]}  
\label{mmm3ps}
\end{equation}
with $s_2$ given by \eq{s2*gen}. 
We see from \eq{mmm3ps} that the boundary metric vanishes except for
$s=s_i$'s, where we have explicitly
\begin{subequations}
\bea
\e^{\varphi(s_1,0)/2}&=&
\frac{1}{\pi \K}\sqrt{\frac{- s t}{s+t}}\frac{s_{42}}{s_{21} s_{41}}, \\
\e^{\varphi(s_2,0)/2}&=&
\frac{1}{\pi \K }\sqrt{\frac{- s t}{s+t}}\frac{s_{31}}{s_{32} s_{21}}, \\
\e^{\varphi(s_3,0)/2}&=&
\frac{1}{\pi \K}
\sqrt{\frac{- s t}{s+t}}\frac{s_{42}}{s_{43} s_{32}}, \\
\e^{\varphi(s_4,0)/2}&=&
\frac{1}{\pi \K }\sqrt{\frac{- s t}{s+t}}\frac{s_{31}}{s_{43} s_{41}}. 
\eea
\label{bm123ps}
\end{subequations}

Calculating the integral with $\varphi$ given by 
\eq{mmm3ps} using the formula
\bea
\lefteqn{\hspace*{-3mm}\int\d y \int_{-\infty}^{+\infty} \d x\,
\frac{y^2}{[(x-s_i)^2+y^2][(x-s_j)^2+y^2]}}\non
&=& \frac{\pi}4 \ln [(s_i-s_j)^2+4y^2],\hspace*{2cm}
\eea
we obtain
\bea
\lefteqn{\hspace*{-3mm}
\K \int_0^\infty \d y \int_{-\infty}^{+\infty} \d x \e^{\varphi(x,y)}}
\non &=&
\alpha'\left(s \ln \frac{s_{43} s_{21}}{s_{42} s_{31}} +
t \ln \frac{s_{32} s_{41}}{s_{42} s_{31}}  \right)
\non &=& \alpha' \left(s \ln \frac s{s+t}+ t \ln \frac t{s+t} \right)
\label{Smmin}
\eea
which reproduces \eq{Smin}. 

Analogously, the length of the boundary contour
equals
\be
\int_{-\infty}^{+\infty}  \d x \e^{\varphi (x,y=0)/2}=0
\ee
as it should for a polygon with light-like edges.

\section{Semiclassical L\"uscher term for the light-like polygon\label{s:Lu}}

\subsection{Semiclassical stringy fluctuations as the L\"uscher term}

The Regge behavior~\rf{Reggestick} with a linear trajectory 
$\alpha(t)=\alpha' t$ of zero
intercept is associated with a classical string. Quantum fluctuations 
shift the intercept of the critical bosonic string to $\alpha(0)=1$.
We shall perform in this Section the computation of the Regge trajectory
for a noncritical string in $d<26$ in the semiclassical approximation.

For a long string, quantum fluctuations can be taken into
account by a semiclassical expansion whose leading order is given
by the minimal area and the semiclassical correction is known
as the L\"uscher term~\cite{LSW80,Lut81}. Its form is explicitly written
for a plane contour via the conformal anomaly: 
\begin{equation}
W(\C) \stackrel{{\rm plane}~\C} \propto  \e^{-\K S_{\rm min}(\C)+
\frac{{d-2}}{96 \pi} \int \d^2 w 
\left(\partial_a \ln\left| \frac{\d z}{\d w}\right|\right)^2 },  
\label{genLuter}
\end{equation}
where an analytic function $w(z)$ maps UHP onto a piece of the plane bounded 
by the contour $\C$.
For a $R\times T$ rectangle with $ {T}\gg {R }$, \eq{genLuter} simplifies to
\begin{equation}
W(\C)
\stackrel{{\rm rectangle}} 
\propto \e^{-\K {RT}+\frac{(d-2)\pi}{24}
{\frac TR} },
\end{equation}
which is more familiar. How the L\"uscher  term emerges for noncritical
strings is demonstrated in Refs.~\cite{Alv81,FTs82,Alv83,DOP84}.
We shall generalize this technique, applying it for the (nonplane) 
momentum-space polygonal loop~\rf{stepx}.

\subsection{Mapping onto rectangle\label{s:3b}}

Let us map the upper half-plane onto a rectangle for arbitrary $s_1$,
$s_2$, $s_3$, $s_4$. By the Schwarz--Christoffel formula we get 
(see \cite{GR}, Eq.~(3.147.4))
\bea
\omega(z)&=& \sqrt{s_{42}s_{31}}\int_{s_2}^z \frac{\d x } 
{\sqrt{(s_4-x)(s_3-x)(x-s_2)(x-s_1)}}\non
&=&2 
F \left( \sqrt{\frac{s_{31}(z-s_2)}{s_{32}(z-s_1)}},
\sqrt{\frac{s_{32} s_{41}}{s_{42} s_{31}}} \right),
\label{SCmap}
\eea
where $F$ is the incomplete elliptic integral of the first kind and
the normalization factor is introduced for the projective symmetry.
The new variable $\omega$ takes values inside a  
rectangle, which has the meaning, as is already said, 
of the worldsheet parametrization. 

Using the relations \cite{GR}, Eq.~(3.147), we find 
\be
R=2C K\left(\sqrt{1-\x}\right),\quad
T=2C K\left(\sqrt{\x}\right),
\ee
where $K$ is the complete elliptic integral of the first kind, 
$C$ is a constant and 
\be
\x=\frac{s_{43} s_{21}}{s_{42} s_{31}}
\ee
is the projective-invariant ratio.
Therefore,
\be
\frac T R = \frac {K\left(\sqrt{\x}\right)}{K\left(\sqrt{1-\x}\right)}
\label{TRx}
\ee
is projective invariant.

To reproduce the mapping of \cite{MO10a}, that corresponds to the choice
$s_1=-1/\sqrt{\mu}$, $s_2=-\sqrt{\mu}$, $s_3=+\sqrt{\mu}$, $s_4=+1/\sqrt{\mu}$, 
we note that
\be
\sqrt{\x}= \frac{1-\mu}{1+\mu},\quad \sqrt{1-\x}= \frac{2\sqrt{\mu}}{1+\mu}.
\ee
Using the formulas \cite{GR}, Eqs.~(8.126.1) and  (8.126.3):
\bea
K \left(\frac{1-\mu}{1+\mu}\right)&=&\frac{1+\mu}2 K(\sqrt{1-\mu^2}),\non
K \left(\frac{2\sqrt{\mu}}{1+\mu}\right)&=&(1+\mu) K(\mu),
\eea 
we then reproduce Eq.~(20) of \cite{MO10a}.

To calculate the L\"uscher term, we decompose
\be
X^\mu(\omega_1,\omega_2)=X^\mu_{\rm cl}
(\omega_1,\omega_2)+Y^\mu_{\rm q}(\omega_1,\omega_2),
\ee
where $X^\mu_{\rm cl}$ is harmonic with the boundary value~\rf{stepx}, 
so $Y^\mu_{\rm q}$ has the mode expansion
\be
Y^\mu_{\rm q}(\omega_1,\omega_2) = \sum_{m,n} \chi^\mu_{m n}
\sin \frac{\pi m \omega_1}R  \sin \frac{\pi n \omega_2}T .
\ee
Now the L\"uscher term results from the determinant coming from the
path integral over $Y^\mu_{\rm q}$.

Using the asymptotes
\be
K\left(\sqrt{\x}\right)\stackrel{\x\to1}\to \frac12 \ln\frac{16}{1-\x},
\qquad K\left(\sqrt{1-\x}\right)\stackrel{\x\to1}\to \frac\pi2, 
\label{asymptotes}
\ee
it is now clear that each set of modes results in the L\"uscher term
\be
\frac{\pi T}{24 R} = \frac 1{24} \ln \frac{16 s}t
\label{TRvsst}
\ee
for $T\gg R$ and
$\x=\x_*=s/(s+t)$. There are $(d-2)$ such sets, so their contribution
to the intercept of the Regge trajectory is
\be
\alpha(0)=\frac{d-2}{24}.
\label{alpha0}
\ee 
It is described in the next section how to get the same result within
the framework of the effective string theory with the action~\rf{effaction}.

\subsection{The effective string theory calculation\label{sec:3c}}

As is already mentioned in Sect.~\ref{ss:Douglas}, 
the way of how the L\"uscher term  emerges for the UHP parametrization
differs from the one for the worldsheet parametrization, described
in the previous Subsection.
It comes now from the classical part $X^\mu_{\rm cl}$ in 
the decomposition~\rf{deco}, rather than from the quantum part $Y^\mu_{\rm q}$.
How this happens for plane contours is described in the 
original paper~\cite{LSW80}, where the determinant of the
Laplace operator in a domain given by the conformal map $w(z)$ was
represented by the integral in the exponent in \eq{genLuter}. 
For this reason the consideration of this Section is pretty much similar
to that of Ref.~\cite{DOP84} for the contribution of the Liouville
field in the Polyakov formulation. This is because the Liouville field 
can be simply substituted to the given order of the semiclassical expansion 
by its value given by the induced metric~\rf{indmetric}.


The conformal symmetry that is maintained in noncritical dimension is
\be
\delta X^\mu = \epsilon (\omega)\partial X^\mu-\frac{\beta a^2}2 \partial^2 
\epsilon (\omega)\frac{\bar \partial X^\mu}{\partial X\cdot \bar \partial X}.
\label{calgebra}
\ee
It transforms $X^\mu$ nonlinearly
--- the same as for the closed string --- and 
the corresponding energy-momentum tensor is
\be
T_{zz} 
=-\frac1{2a^2} \partial X\cdot \partial X +\frac{\beta}2 
\frac{\partial^3 X \cdot \bar\partial X }{\partial X \cdot \bar\partial X }
\ee
with $\K=1/4\pi a^2$, so that $2a^2=\alpha'$.
Expanding around the classical solution
\be
X_{\rm cl}^1= \omega_1 \sqrt{RT},\quad X_{\rm cl}^2= \omega_2 \sqrt{RT},
\quad X_{\rm cl}^0=X_{\rm cl}^3=0
\ee
or
\be
X_{\rm cl}^\mu= \left(e^\mu \omega +\bar e^\mu \bar\omega \right) \sqrt{RT}
\label{Xcl}
\ee
with
\be
e^\mu=\left( 0,{\frac 12}, -{\frac {\i} 2},0 \right),
\ee
where $\omega$ takes values inside a
$\sqrt{R/T}\times \sqrt{T/R}$ rectangle,
we obtain
\be
T_{zz}=
-\frac{\sqrt{RT}}{a^2} e\cdot \partial Y_{\rm q}
-\frac{1}{2a^2}\partial Y_{\rm q} \cdot\partial Y _{\rm q}+
\frac{\beta}{\sqrt{RT}} \bar e \cdot \partial^3 Y_{\rm q}. 
\label{T--}
\ee
Using the Dirichlet Green function for UHP
\be
G(z,\zeta)=-\frac1{2\pi\K} \ln \left| \frac{z-\zeta}{z-\bar\zeta}
\right|,
\label{56}
\ee
we get for the rectangle by the (inverse) conformal mapping~\rf{SCmap}:
\be
G(\omega,\Omega)=-\frac1{2\pi\K} \ln \left| 
\frac{\sn^2 \frac{\omega}2 -\sn^2 \frac{\Omega}2}
{\sn^2 \frac{\omega}2 -\overline{\sn^2 \frac{\Omega}2}}
\right|
\label{Greenomega}
\ee
with $\sn\,\alpha\equiv \sn (\alpha, \sqrt{1-r})$ being the Jacobi elliptic
function.

Equations~\rf{Xcl} and \rf{T--} look like those of Ref.~\cite{PS91} 
for a winding closed string with
$R$ replaced by $\sqrt{RT}$, so we repeat the computation of the central
charge generated by the conformal transformation~\rf{calgebra} to
obtain analogously 
\be
\LA T_{zz}(\omega) T_{zz}(\Omega) \RA_Y = 
\frac{d+ 12 \beta}{2(\omega-\Omega)^4} 
+{\cal O}\left((\omega-\Omega)^{-2}\right), 
\ee
where the averaging over the fluctuating field $Y^\mu_{\rm q}$ is given by
the Green function~\rf{Greenomega}. This fixes 
\be
\beta = \frac{26-d}{12}
\ee
in our case of an open string as well.
 
Similar formulas can be obtained for the UHP parametrization,
when $\omega(z)$ in \eq{Xcl} is given by the mapping~\rf{SCmap}.
Now it should be noted that the induced metric
\be
\e^{\varphi}\equiv 2 \partial X_{\rm cl} \cdot \bar\partial X_{\rm cl} 
=2 RT \frac{s_{42}s_{31}}
{\prod_{i=1}^4 \sqrt{(x-s_i)^2+y^2}} 
\ee 
is not constant, as it is for the worldsheet parametrization.
For a general choice of $s_1$, $s_2$, $s_3$, $s_4$ we have
\be
\frac{\partial^2 X_{\rm cl} \cdot \bar\partial^2 X_{\rm cl}}
{\partial X_{\rm cl} \cdot \bar \partial X_{\rm cl}} 
=\partial \varphi\bar\partial \varphi
+\partial \bar \partial \varphi
\ee
and
\bea
\lefteqn{\hspace*{-3mm}\int_0^\infty \d y \int_{-\infty}^{+\infty} \d x\,
(\partial_a \varphi)^2}\non &=&
-\frac \pi 2\sum_{i,j=1}^4 \ln \left[(s_i-s_j)^2
+(\varepsilon_i+\varepsilon_j)^2 \right].
\eea
For $(s_{i}-s_{i-1})\gg \varepsilon_i,\varepsilon_{i-1}$ this gives
\bea
\hspace*{4mm}
\lefteqn{\hspace*{-5mm}\frac{(d-26)}{96\pi}
\int_0^\infty \d y \int_{-\infty}^{+\infty} \d x\, 
(\partial_a \varphi)^2}\non
&=&-\frac{(d-26)}{96}\ln \left[16 s_{43}^2 s_{42}^2 s_{41}^2 s_{32}^2 s_{31}^2  
s_{21}^2\varepsilon_1\varepsilon_2\varepsilon_3\varepsilon_4 \right] \non
&=&-\frac{(d-26)}{24}\ln \left[2 s_{43} s_{32} s_{21} s_{41}\varepsilon \right]
\label{ssss}
\eea
provided
\be
\varepsilon_i=\frac{(s_{i+1}-s_{i})(s_{i}-s_{i-1})}{(s_{i+1}-s_{i-1})}
\varepsilon
\label{regueps}
\ee
as is prescribed by the covariance, where $\varepsilon$ is an invariant cutoff 
of the dimension of length.
Equation~\rf{ssss} now reproduces the L\"uscher term as $\x\to1$:
\bea
\lefteqn{\hspace*{-3mm}\frac{(d-26)}{96\pi}
\int_0^\infty \d y \int_{-\infty}^{+\infty} \d x\, 
(\partial_a \varphi)^2}\non
&\to &-\frac{(d-26)}{24}\ln \left[2(1-\x) \varepsilon \right],
\label{inview}
\eea
in view of Eqs.~\rf{TRx}, \rf{asymptotes}.

Actually, this
calculation repeats the one of Ref.~\cite{DOP84} for the Polyakov string
because there is apparently no difference between the induced and intrinsic 
metrics through this order of the semiclassical expansion. 
Alternatively, in the worldsheet parametrization the L\"uscher term comes
from the determinant resulting from the path integration over $Y^\mu_{\rm q}$.
As was originally pointed out in Ref.~\cite{LSW80}, this determinant
equals precisely the left-hand side of \eq{inview}.
We have thus illustrated the statement already made in Sect.~\ref{ss:Douglas}
concerning the difference between the UHP and worldsheet parametrizations. 

We are now in a position to compute a semiclassical correction
to the Regge trajectory of the effective string theory in $d< 26$.
Using \eq{TRvsst} and substituting $\alpha(0)=1$ for the 
intercept of the critical string, we obtain
\be
\alpha(0)=1+ \frac{d-26}{24}=\frac{d-2}{24},
\ee
reproducing \eq{alpha0}. 
It worth emphasizing that the expansion of the effective string theory 
goes for the scattering amplitude in the parameter
\be
\left(\ln\frac1{1-r}\right)^{-1}=\left(\ln\frac st\right)^{-1},
\label{67}
\ee 
like it was $R^{-1}$ in Ref.~\cite{PS91}. Therefore, the expansion is 
justified by the Regge kinematical regime
and we assume that the semiclassical Regge trajectory~\rf{linea}
may turn out to be exact.

\section{Generalization to $\Delta p_i^2\neq0$\label{s:neq}} 

If $\Delta p_i^2 \neq 0$, we have to keep the term with $j=i$
in the above equations. Then \eq{DouminM} is replaced by
\be
-\sum_{j\neq i} \frac{2\Delta p_i \cdot \Delta p_j}{s_i-s_j}
+\pi \sum_j \Delta p_j^2 
\left\langle\frac{\partial G\left(s_j,s_j\right)}{\partial s_i}
\right\rangle=0,
\label{DouminMii}
\ee
where
\be
\left\langle G\left(s_j,s_j\right)\right\rangle
=\frac{\int \D_{\rm diff} s \,G\left(s_j,s_j\right)}{\int \D_{\rm diff} s}
\ee
with the reparametrization path integral going over functions obeying
$s(t_i)=s_i$ which are zero modes of the Douglas minimization.

Substituting
\be
\left\langle G\left(s_j,s_j\right)\right\rangle=\frac{1}{\pi}
\ln \frac{(s_{j+1}-s_{j-1})}{(s_{j+1}-s_{j})(s_{j}-s_{j-1})\varepsilon},
\ee
that corresponds to the Lovelace choice as is discussed 
in Refs.~\cite{DiV86,MO10b}, we get
\bea
\hspace*{-3mm}
&&-\sum_{j\neq i} \frac{2\Delta p_i \cdot \Delta p_j}{s_i-s_j}+\Delta p_{i-1}^2 
\left[ \frac{1}{(s_i-s_{i-2})}-\frac{1}{(s_i-s_{i-1})}\right]
\hspace*{-2mm}\non
&&-
\Delta p_{i}^2 \left[ \frac{1}{(s_i-s_{i-1})}-\frac{1}{(s_{i+1}-s_{i})}\right]
\non &&+\Delta p_{i+1}^2
\left[\frac{1}{(s_{i+1}-s_{i})}-\frac{1}{(s_{i+2}-s_{i})}\right]
=0.
\label{DouminMiiL}
\eea
For $M=4$ this results in the same formula \rf{s2*gen} with 
$\Delta p_i^2$ included in the definition of the Mandelstam variables.

For $\Delta p_i^2\neq 0$ the term 
\be
\frac1{\pi^2\K^2} \sum_{i} 
\frac{\Delta p_i^2\, y^2}{[(x-s_i)^2+y^2]^2}
\label{mmmii}
\ee
appears additionally in the induced metric. 
It is more singular at the boundary than~\rf{bm123ps}, resulting in
\bea
\e^{\varphi(s,0)/2}
&= &\sum_i \frac{\sqrt{\Delta p_i^2}}{\pi \K} 
\frac{\varepsilon_i}{[(s-s_i)^2+\varepsilon_i^2]}\non
&\to &
 \sum_i \frac{\sqrt{\Delta p_i^2}}{\K} 
\delta(s-s_i). 
\eea
This reproduces $\sqrt{\Delta p_i^2}/K$ for the lengths of the polygon edges.


It still has to be verified, however, whether or not the conformal gauge 
is maintained by this construction for $\Delta p_i^2\neq 0$.

\section{Application to QCD}

As is already mentioned, QCD string is stretched between quarks, when
they are separated by large distances. The results of Refs.~\cite{MO08,MO10b}, 
which state that large loops dominate the sum-over-path representation
of QCD scattering amplitudes in the Regge kinematical regime, assume
therefore an applicability of the effective string theory ideology 
in this case.

To illustrate this issue, we start from the representation of $M$-particle 
scattering amplitudes in large-$N$ QCD through the Wilson loops:
\begin{eqnarray}
&&\hspace*{-4mm}
A\left(\Delta p_1,\ldots, \Delta p_M \right)\propto
\int\nolimits_0^\infty \d \Tau\, \Tau^{M-1} \e^{-m \Tau} \non &&\times
\int\nolimits_{-\infty}^{+\infty} \frac{\d t_{M-1}}{1+t_{M-1}^2}
\prod_{i=1}^{M\!-\!2}\int\nolimits_{-\infty}^{t_{i\!+\!1}} \frac{\d t_i}{1+t_i^2} 
\non &&
\times  \hspace*{-8mm}
\int\limits_{x^\mu(-\infty)=x^\mu(+\infty)=0} \hspace*{-8mm} \D x^\mu(t) \,
\e^{\i \int \d \tau\, \dot x (t)\cdot p(t)}\,
J[x(t)] \, W[x(t)], ~ 
\label{117}
\end{eqnarray}
where $p^\mu(t)$ is the piecewise constant momentum-space loop~\rf{stepx}.
For spinor quarks and scalar operators, the weight for the path integration 
in \eq{117} is
\be
J[x(t)]=\int \D k^\mu(t)\;{\rm sp\;} \boldsymbol{P}
\e^{\i \int \d t\,
[\dot x (t)\cdot k(t)- \Tau \gamma\cdot k(t)/(1+t^2)]} ,
\label{J}
\ee
where \/sp \/and the path-ordering refer to $\gamma$-matrices.
In \eq{117} $W[x(t)]$ is the Wilson loop in pure Yang--Mills theory at large $N$
(or quenched), $m$ is the quark mass and $\Tau$ is the proper time.
For finite $N$, correlators of several Wilson loops would have to be taken into 
account.

For the critical string, when Eqs.~\rf{ansatz} and \rf{Di} are expected to hold 
for large contours, the path integral over $x^\mu(t)$ in \eq{117} is Gaussian 
and can be done as is outlined in Refs.~\cite{MO08,MO10b}. It is saturated in
that case by 
the classical trajectory
\bea
x^\mu_{\rm cl}(t)&=&\i \alpha' \int_{-\infty}^{+\infty}\d t'\, \dot p^\mu(t')
\ln{[s(t)-s(t')]^2}\non &=&\i \alpha'\sum_j \Delta p_j^\mu \ln{[s(t)-s_j]^2},
\label{clax} 
\eea
which is $T$-dual in the sense of 
Ref.~\cite{AM07a} to that given by \eq{stepx}, giving the same magnitude
of the minimal area. It is pure imaginary like this often happens for
a saddle point of integrals with an oscillating integrand.
Doing the integral over $x^\mu(t)$, we finally obtain
\begin{eqnarray}
&&\hspace{-4mm} A\left(\Delta p_1,\ldots, \Delta p_M \right) \propto
\int\nolimits_0^\infty \d \Tau\, \Tau^{M-1} \e^{-m \Tau}  \non &&\times
\int\nolimits_{-\infty}^{+\infty} \frac{\d t_{M-1}}{1+t_{M-1}^2}
\prod_{i=1}^{M\!-\!2}\int\nolimits_{-\infty}^{t_{i\!+\!1}} \frac{\d t_i}{1+t_i^2}\non 
&& \times \int \D k^\mu(t)\;{\rm sp\;} \boldsymbol{P}
\e^{-\i\Tau  \int \d t\,\gamma\cdot k(t)/(1+t^2)} \non &&\times\,
W\Big[x_*(t)=\frac1K \left(p(t)+k(t)\right)  \Big].
\label{118}
\end{eqnarray}

For $d\leq 26$ we substitute the Wilson loop in the form of 
the disk amplitude for the effective string theory with 
the action~\rf{effaction}:
\begin{equation}
W[x(\cdot)]=\int {\cal D}_{\rm diff} t(s)\!\!\!\!\!
\int\limits_ {X^\mu(x=s,0)=x^\mu(t(s))}\!\!\!\!\!
{\cal D} X^\mu(x,y)\e^{-\K S_{\rm eff}},
\label{effansatz}
\end{equation}
which reproduces Eqs.~\rf{ansatz} and \rf{Di} in $d=26$, when $S_{\rm eff}$
is quadratic in $X^\mu$.
The path integral over $x^\mu(t)$ in \eq{117} can also be done for $d<26$
within the semiclassical expansion. Equation~\rf{clax} is then modified in
the semiclassical approximation as
\begin{equation}
x^\mu(t)=x^\mu_{\rm cl}(t)+
 \alpha' \int_{-\infty}^{+\infty}\d t'\, 
\ln{[s(t)-s(t')]^2}\,\frac{\delta S_{\rm eff}^{(2)}}{\delta x_\mu (t')},
\label{claxx} 
\end{equation}
where  $S_{\rm eff}^{(2)}$ stands for the second term on the right-hand side of 
\eq{effaction}.
Equation~\rf{118} remains unchanged with this accuracy. 
Details of the derivation are described in Appendix~\ref{appB}.

As distinct from its stringy counterpart~\rf{Psip},
the right-hand side of \eq{118} has the additional path integration over 
$k^\mu(t)$,
which emerges from Feynman's disentangling of the $\gamma$-matrices.
For small $m$ and/or very large $M$, the integral over $\Tau$ in \eq{117} 
is dominated by large $\Tau\sim (M-1)/m$. 
Then typical values of $k\sim 1/\Tau$ are essential for large $\Tau$ 
in the path integral over $k^\mu(t)$, and 
we can disregard $k(t)$ in the argument of $W$ in \eq{118},
so the path integral over $k^\mu(t)$ factorizes. 
We finally obtain~\cite{MO08} from \eq{117} the product of the string scattering
amplitude $A\left[p(t)\right]$ times factors which do not depend on $p$.
The substitution of the effective string theory representation~\rf{effansatz} 
into \eq{117} for $d<26$ 
results in a more complicated path integral over $x^\mu(t)$ which is, however,
Gaussian within the semiclassical expansion, reproducing again \eq{118}.

Thus, the scattering amplitude $A\left(\{\Delta p_i\}\right)$ 
coincides for the ansatz~\rf{effansatz}
with $W\left[x_*(t)\right]$, where $x^\mu_*(t)$ is given by \eq{stepx}
and the reparametrization path integral goes over the functions $s(t)$, 
obeying $s(t_i)=t_i$. 
Therefore, \eq{118} reproduces for piecewise constant $p^\mu(t)$
the Regge behavior of (off-shell) scattering amplitudes 
in the effective string theory
as $m\rightarrow0$ and/or $M\to\infty$.
Since we are dealing with the disk amplitude, associated with 
planar diagrams, we identify this Regge trajectory
with the quark-antiquark Regge trajectory%
\footnote{It is often called also as the Reggeon or the secondary Regge
trajectory 
to be distinguished from the vacuum Regge trajectory = Pomeron.} 
in large-$N$ QCD and thus conclude that it is linear 
in the semiclassical approximation, while the actual intercept 
can be larger than the value given by \eq{linea} owing to the breaking 
of the chiral symmetry, as is pointed out in Ref.~\cite{MO10b}.
The linear trajectory seems to disagree with the old 
results~\cite{FTs82,FTs83}, where the path integral over the Liouville 
field was Gaussian with either Neumann or Dirichlet boundary conditions, 
so zero modes associated with reparametrizations 
were not taken into account, as it is done now. To my understanding,  
this emphasizes the very important role played by the reparametrization 
path integral.


\section{Conclusion}

We have shown in this Paper that the Regge asymptote of scattering
amplitudes can be obtained within the ideology of an effective string
theory and is not affected by short distances.
For this reason, these results are also applicable to QCD string
which is generically not the Nambu--Goto one, but behaves like it at
large distances. 
The expansion goes around a long-string configuration and has the
meaning of a semiclassical expansion, whose parameter $1/\ln(s/t)$ is small 
in the Regge kinematical regime of $s\gg-t$.

A linear Regge trajectory~\rf{linea} 
of a noncritical string had been vastly discussed in the literature.%
\footnote{For a historical review see Ref.~\cite{DiV08}.}
The intercept $(d-2)/24$
is precisely the value which follows from the spectrum~\cite{Arv83}. 
This result is most probably consistent because the anomaly emerging in
the Virasoro quantization vanishes for long strings, as was
pointed out in Ref.~\cite{Ole85}.

It is interesting to discuss the relation between our results on
the Regge behavior of QCD scattering amplitudes
in the framework of the effective string theory and
similar known results on the Pomeron~\cite{JP00,Jan01} and 
the Reggeon~\cite{JP02} (the one we consider) trajectories 
in the framework of the AdS/CFT correspondence in a confining background.
While the minimal surfaces describing the classical part are constructed
in both cases for a flat metric, they are apparently different because
Refs.~\cite{JP00,Jan01,JP02} use an impact-parameter representation
of the scattering process
and this Paper deals with polygonal loops in momentum space. 
An advantage of our approach is the existence of
a systematic expansion in the parameter~\rf{67}.
The way we have calculated the intercept in Sect.~\ref{s:3b} via semiclassical 
fluctuations of the minimal surface (given by the L\"uscher term) is
pretty much similar to the one in Refs.~\cite{Jan01,JP02} except for
the difference in the number of fluctuating transverse degrees of freedom,
that equals 2 in our case for $d=4$ from the consistency 
of the effective string theory in hand, 
which is also favored by the lattice simulations~\cite{Tep09,ABT10}
as is already mentioned in the Introduction.
This issue can be further clarified by extending our calculations to
an annulus amplitude which is to be associated with the Pomeron exchange.

It is also worth mentioning that a semiclassical
calculation of the intercept in the framework of
the effective string theory, which is close spiritually to our
calculation, was performed in Ref.~\cite{BS02}
from the spectrum of a rotating string.
We emphasize once again that the consideration of this Paper refers
to the scattering domain of $t<0$ and deals with the scattering amplitudes.
  

As distinct from the Polyakov formulation, where the intrinsic metric is
treated as an independent variable, in
the effective string theory the worldsheet metric is induced.
The path integration now goes only over the embedding-space coordinate
$X^\mu$ and reparametrizations of the boundary contour. The former path
integral turns out to be Gaussian within the semiclassical expansion,
while the latter one has a well-defined measure and has been recently 
studied both analytically~\cite{MO08} and numerically~\cite{BM09}. 
Therefore, the issue of integrating over the Liouville field in the 
bulk, which was the subject of Ref.~\cite{DDK}, does not emerge.
It would be very interesting to calculate~\cite{AM11} the string susceptibility
$\gamma_{\rm str}$ for large areas within the effective string theory 
approach and to compare with the existing results.
 
\begin{acknowledgments}
I am indebted to Poul Olesen for invaluable discussions and suggestions.
I am also grateful to Paolo Di Vecchia  
and Pawel Caputa for useful discussions, and Romuald Janik for
the email correspondence.
\end{acknowledgments}

\appendix

\section{Proof of the conformal gauge for Douglas' minimization\label{appA}}

Let us show that $X^\mu(x,y)$ obeys the conformal gauge
\begin{subequations}
\bea
\partial_x X\cdot \partial_y X &=&0,
\label{conf1}\\*
\partial_x X\cdot \partial_x X &=&\partial_y X\cdot \partial_y X ,
\label{conf2}
\eea
\label{conf}
\end{subequations} 
if the Douglas minimization~\rf{Doumin} is imposed on $s(t)$ for
the given boundary contour $x^\mu(t)$.

We substitute
\be
\partial_y \frac{y}{(x-s)^2+y^2}=-\partial_x \frac{(x-s)}{(x-s)^2+y^2},
\ee
integrate by parts and use the following two identities
\begin{widetext}
\begin{subequations}
\bea
\hspace*{-6mm}
\frac1{(s_1-s_2)}&&\hspace*{-3mm}\left\{\frac1{[(x-s_1)^2+y^2]}
-\frac1{[(x-s_2)^2+y^2]}\right\}=
\frac{(2x-s_1-s_2)}{[(x-s_1)^2+y^2][(x-s_2)^2+y^2]},
\label{iden1}\\ \hspace*{-6mm}
\frac1{(s_1-s_2)}&&\hspace*{-3mm}\left\{\frac{(x-s_2)}{[(x-s_1)^2+y^2]}
-\frac{(x-s_1)}{[(x-s_2)^2+y^2]}\right\}
=\frac{y^2+(x-s_1)^2+(x-s_2)^2+(x-s_1)(x-s_2)}{[(x-s_1)^2+y^2][(x-s_2)^2+y^2]}
\label{iden2}
\eea
\label{iden}
\end{subequations} 
for Eqs.~\rf{conf1} and \rf{conf2}, respectively.
They are then satisfied if the minimization equation \rf{Doumin} is fulfilled.

For \eq{conf1} we have
\bea
\partial_x X\cdot \partial_y X &= &-
\int \frac{\d t_1}{\pi} \frac{\d t_2}{\pi}
\frac{(x-s(t_1))y}{[(x-s(t_1))^2+y^2][(x-s(t_2))^2+y^2]} \non
 &= &-
\int \frac{\d t_1}{\pi} \frac{\d t_2}{\pi}
\frac{(2x-s(t_1)-s(t_2))y}{2[(x-s(t_1))^2+y^2][(x-s(t_2))^2+y^2]}
=0
\eea
in view of \eq{iden1} and \eq{Doumin}.

For \eq{conf2} we have
\bea
\partial_x X\cdot \partial_x X -\partial_y X\cdot \partial_y X &= &
\int \frac{\d t_1}{\pi} \frac{\d t_2}{\pi}
\frac{(x-s(t_1))(x-s(t_2))-y^2}{[(x-s(t_1))^2+y^2][(x-s(t_2))^2+y^2]}
\non &= &
\int \frac{\d t_1}{\pi} \frac{\d t_2}{\pi}
\frac{(x-s(t_1))(x-s(t_2))+y^2+(x-s(t_1))^2+(x-s(t_2))^2}
{[(x-s(t_1))^2+y^2][(x-s(t_2))^2+y^2]}
=0
\eea
\end{widetext}
in view of \eq{iden2} and \eq{Doumin}.

\section{Derivation of \eq{118}\label{appB}}

We collect in this Appendix some formulas which are used in the derivation
of \eq{118}.

First of all, let us explain how to understand the variational derivative
in \eq{claxx}. The point is that $S^{(2)}_{\rm eff}$ is a functional of
the bulk variable $X^\mu(x,y)$, while $x^\mu(t(s))$ is its boundary value.
At the classical level, \eq{harmox} holds and we have
\be
\frac{\delta X^\mu_{\rm cl}(x,y)}{\delta x^\nu (t)}= \delta^\mu_\nu \frac {\dot s(t)}\pi
\frac{y}{(x-s(t))^2+y^2},
\ee
which reproduces the standard delta-function as $y\to0$. 

To derive \eq{118}, it is convenient to use the short-hand notation
\be
G*f(t)\equiv \int \d t'\, G\left(s(t)-s(t')\right) f(t').
\ee
Then, for example, \eq{clax} takes the form
\be
x^\mu_{\rm cl}(t)= -\frac \i \K G*\dot p^\mu(t)
\label{newclax}
\ee
with
\be
G(s)=-\frac 1\pi \ln |s|.
\ee

For the exponent in the path integral we have
\be
\i p_\mu*\dot x^\mu-S_{\rm eff}[X]
\label{expom}
\ee 
whose Euler--Lagrange equation including a semiclassical correction reads
\be
-\i \dot p_\mu - \K G^{-1}*x_\mu-
\frac{\delta S^{(2)}_{\rm eff}}{\delta x^\mu}=0,
\label{B6}
\ee
where the inverse to $G$ is
\be
 G^{-1}\left(s(t)-s(t')\right)=\frac{\d}{\d t} \frac{\d}{\d t'} 
 G\left(s(t)-s(t')\right). 
\ee
An iterative solution to \eq{B6} is given by \eq{claxx}.
Substituting this into the exponent \rf{expom},
we finally obtain to the given order of the semiclassical expansion
\be
\rf{expom}=-\frac 1{2\K}\dot p_\mu*G *\dot p^\mu -S^{(2)}_{\rm eff}[X_{\rm cl}],
\ee
where $X^\mu_{\rm cl}$ is reconstructed from \rf{clax}  by \eq{harmox}.

The last step in proving \eq{118} is to note that $X^\mu_{\rm cl}$, reconstructed
from the boundary value \rf{clax}, can be replaced by \rf{amostX},
reconstructed from the boundary value \rf{stepx}. The point is that
$\ln\sqrt{(x-s_i)^2+y^2}$ and $\arctan[(x-s_i)/y]$ are real and imaginary parts
of an analytic function $\ln(z-s_i)$ and obey the Cauchy--Riemann equations.
Therefore, $S_{\rm eff}$ does not change under such a replacement. This is
similar to the $T$-duality transformation in Ref.~\cite{AM07a}. 

We have thus proved \eq{118}.


\end{document}